\documentclass[11pt]{article}
\usepackage[utf8]{inputenc}
\usepackage{graphicx}
\usepackage{authblk} 
\usepackage[margin=1in]{geometry} 
\usepackage{mathptmx} 
\usepackage[round]{natbib} 
\usepackage{array} 
\usepackage{float}

\usepackage[colorlinks=true, urlcolor=blue, linkcolor=blue, citecolor=blue]{hyperref}

\title{\textbf{The Need for Benchmarks to Advance AI-Enabled Player Risk Detection in Gambling}}

\author[1]{Kasra Ghaharian\thanks{Corresponding author: kasra.ghaharian@unlv.edu}}
\author[1]{Simo Dragicevic}
\author[1]{Chris Percy}
\author[2]{Sarah E. Nelson}
\author[3]{W. Spencer Murch}
\author[4]{Robert M. Heirene}
\author[1,5]{Kahlil Simeon-Rose}
\author[6]{Tracy Schrans}

\affil[1]{\small International Gaming Institute, University of Nevada, Las Vegas, USA}
\affil[2]{\small Division on Addiction, Cambridge Health Alliance, Harvard Medical School, USA}
\affil[3]{\small Department of Psychology, University of Calgary, Canada}
\affil[4]{\small Brain \& Mind Centre, School of Psychology, Science Faculty, University of Sydney, Australia}
\affil[5]{\small Carson College of Business, Washington State University, Everett, USA}
\affil[6]{\small Focal Research Consultants, Canada}

\date{}

\begin{document}

\maketitle

\begin{abstract}
\noindent %
Artificial intelligence-based systems for player risk detection have become central to harm prevention efforts in the gambling industry. However, growing concerns around transparency and effectiveness have highlighted the absence of standardized methods for evaluating the quality and impact of these tools. This makes it impossible to gauge true progress; even as new systems are developed, their comparative effectiveness remains unknown. We argue the critical next innovation is developing a framework to measure these systems. This paper proposes a conceptual benchmarking framework to support the systematic evaluation of player risk detection systems. Benchmarking, in this context, refers to the structured and repeatable assessment of artificial intelligence models using standardized datasets, clearly defined tasks, and agreed-upon performance metrics. The goal is to enable objective, comparable, and longitudinal evaluation of player risk detection systems. We present a domain-specific framework for benchmarking that addresses the unique challenges of player risk detection in gambling and supports key stakeholders, including researchers, operators, vendors, and regulators. By enhancing transparency and improving system effectiveness, this framework aims to advance innovation and promote responsible artificial intelligence adoption in gambling harm prevention.

\vspace{1em} 
\end{abstract}


\section{Introduction}
Gambling is a significant and growing form of entertainment, participated in by millions globally. While most consumers enjoy gambling as a leisure activity, prevalence rates for harm vary considerably across different populations, jurisdictions, and product types. Globally, an estimated 1.4\% of adults meet criteria for problem gambling, but these rates shift dramatically depending on the gambling environment; for example, prevalence rises to approximately 16\% among online casino or slots gamblers \citep{Tran2024}. The industry’s rapid expansion, driven largely by digitization and the global availability of online and mobile products, has led to a corresponding increase in efforts to mitigate these potential harms \citep{Ghaharian2022}. These collective efforts are commonly referred to as Responsible Gambling (RG)\footnote{We use the term Responsible Gambling (RG) throughout this paper as it is commonly used to describe these practices. We acknowledge the ongoing and important discourse contrasting the RG framework to safer gambling or public health approaches, but a full discussion of that debate is beyond the scope of this paper.}.

RG strategies have evolved as a direct result of gambling’s digitization, and include targeted RG messaging, immediate access to gambling help resources, and automated risk detection systems \citep{Shaffer2020}. These technological player protection measures can be divided roughly into two types: user-initiated measures and automated operator-initiated strategies. The former includes tools that users can employ, usually voluntarily, to control their gambling: self-limiting features, reality checks, deposit limits, cool-off periods, and self-exclusion. However, few users take advantage of these features \citep{Ladouceur2016, Nelson2008}, and there is debate about whether those who do are the players who need them most \citep{catania2021understanding}. 

While automated, operator-initiated strategies can also take many forms, including universal approaches like deposit limits, there has been a movement in the past decade toward more tailored interventions \citep{heirene2025setting}. Universal strategies risk either disrupting the activity of those who do not need protections or, if implemented in a way that does not disrupt most users’ activity, being set so conservatively that many of those who need them do not trigger them \citep{broda2008optimal}. On the other hand, targeted strategies can, in theory, use player information to better determine when and to whom to deploy protections.

As such, data-driven RG models have become the holy grail of automated operator-initiated RG solutions \citep{Ghaharian2022}. These systems often leverage artificial intelligence (AI) – more specifically, machine learning (ML) – to analyze players’ gambling behavior and detect patterns associated with harm to trigger interventions. This approach has attracted substantial interest from both academia and industry stakeholders, with several regulators mandating the use of data-driven systems for player risk detection, operators developing proprietary models, and specialist technology vendors emerging to offer commercial solutions.

However, a fundamental question remains largely unanswered: \textit{Are these sophisticated systems effective?}

This paper argues that the next critical innovation required is not the advancement of current methods or the creation of new models, but the development of a framework to measure them. We propose that this problem can be addressed by applying Performance Benchmarking, a common practice in many AI domains, but one that is yet to be applied in this use case.

The paper is presented as follows: We first provide a brief background on the use of ML for player risk detection, followed by a discussion of the current challenges in evaluating these systems. We then introduce the concept of Performance Benchmarking and briefly review its application in other domains. Finally, we present a conceptual, discussion-starting framework for a player risk detection benchmarking suite. Rather than proposing a single, fully specified benchmark, we outline the structural requirements, core dimensions, and practical considerations necessary for such benchmarks to be developed and adopted by the research and practitioner communities. Our hope is to stimulate discussion and thought for what such a framework would require and what it could achieve.

\subsection{Machine Learning for Player Risk Detection}

A ML approach to player risk detection typically uses information about gambling patterns and player characteristics from gambling records (behavioral markers) to forecast potential gambling-related problems\footnote{For an up-to-date and comprehensive review of model architectures and the current landscape, we refer the reader to \citealp{MARIONNEAU2025100644}}. This is similar to the use of biomarkers in medicine, where objectively measured characteristics serve as indicators of disease processes \citep{biomarkers2001biomarkers}. Medicine also uses behavioral markers, for example, in early detection of Alzheimer’s disease \citep{gomar2011utility}.

ML models can target players according to markers defined by both their gambling patterns and changes in those patterns, using this information to designate differing levels of potential risk. Most importantly, these models can provide a guide for matching responses to individuals, optimizing the opportunity for favorable outcomes. Very basic best practices for developing such models include establishing that the markers differ in the groups to be distinguished (e.g., individuals who exhibit problematic gambling behavior and those who gamble without problems), developing a predictive model that integrates markers to predict the outcome in question, and then establishing the model’s performance in samples separate from those used to develop the model \citep{Morrow2008, vanderstichele2008}.

One distinct feature of ML is its ability to learn complex, nonlinear relationships in data, making it highly suited for identifying at-risk play given the individual variability in gambling behavior \citep{hastie2009elements}. However, complex ML solutions encounter some of the same challenges as the most basic models, such as establishing accurate outcome measures and training data and developing models that can generalize and replicate beyond the idiosyncrasies of a specific dataset (i.e., distinguishing the signal from the noise: \citealp{silver2012signal}; also see \citealp{Shaffer2020}).

Early player risk models re-purposed data from loyalty smart cards used in physical gambling venues (e.g., casinos) \citep{Schellinck2011}. The shift to online gambling, however, significantly simplified behavioral tracking. Unlike physical venues that depended on a player inserting a card, online platforms and mobile apps inherently capture comprehensive player data. This seamless data collection fueled a new research area, leading to at least 40 academic studies and several literature reviews (we refer the interested reader to: \citealp{Ghaharian2022,Delfabbro2023,MARIONNEAU2025100644}). Alongside this academic work, commercial models were developed. A recent review by \citet{MARIONNEAU2025100644} identified 11 commercial models, including six in-house systems from operators (e.g., Kindred, Veikkaus) and five from third-party vendors (e.g., Mindway, Openbet).

These models vary widely in their methods, employing a variety of unsupervised and supervised algorithms, and in the number and type of input markers they use. As \citet{MARIONNEAU2025100644} note, these markers are often unclear or inconsistently reported in the literature, and the number of markers used in a given model range from fewer than 10 to over 100. \citet{ghaharian2025ai} recently mapped the markers in the academic literature, identifying 65 “low-level” categories (such as bet amount and net loss) which can be grouped into five “high-level” categories: play (betting behavior), payment (financial transactions, i.e., deposits and withdrawals), engagement (when, how often, and how broadly players engage with gambling), RG tool use, and profile information (e.g., registration date, demographic information).

While literature reviews have helped map the academic literature, comparing model performance remains very difficult. Studies often use different datasets, model parameters, and data cleaning procedures, making it hard to draw clear conclusions about which methods are most effective. This challenge is compounded by the significant transparency problem with commercial models, as the information available on these systems is insufficient for a meaningful comparison. These and other challenges are discussed in the following section.

\section{Current Problems}

It is appropriate to hold ML to high standards, given that policy increasingly references it in supporting harm reduction \citep{bedford2024high}. Moreover, such standards align with the sensible and realistic performance beliefs that gamblers themselves may hold \citep{murch2025preferences}. Several challenges limit the potential for ML in this context.

\subsection{Broad Challenges for Stakeholders}

ML, as with all identification and intervention design, involves trade-offs between sensitivity and precision, i.e., where to place the balance between failing to identify gamblers at risk and falsely identifying many gamblers with negligible risk levels \citep{Murch2023}. These trade-offs can be improved with more sophisticated models but cannot be fully resolved in real-world settings. In the absence of precision, an ultra-low specificity model simply provides RG interventions to all players, which may undermine trust in RG programs and potentially lead to substitution by players towards other operators or products.

As models grow more complex, they typically become less transparent with respect to how any individual decision has been reached \citep{sarkar2016accuracy}. This is often referred to as the “black box” problem, where the model’s internal logic is so intricate that it becomes opaque even to its creators, making it difficult to audit or explain a specific outcome \citep{vonEschenbach2021}. Explainability, or the ability to identify how a model arrives at its prediction, is important to facilitate users’ trust in a system and ensure the model is operating as expected \citep{sarkar2016accuracy}. In addition, explainability has practical relevance: it enables operators to activate appropriate interventions and allows players to reflect on behavioral dimensions that have been identified as increasing their risk.

Another issue is that models can only identify patterns in the data they possess. Consequently, identical gambling behaviors can represent very different risk-levels depending on variables not included in a data set, such as an individual player’s life circumstances or financial context. For example, the potential impact of a specific monetary loss is difficult to contextualize without information about an individual’s financial circumstances (e.g., income, debt exposure, etc.). Moreover, harms are not purely financial. A significant amount of time spent gambling could be a harmless hobby for one person, but for someone suffering psychological distress, it could be a coping mechanism leading to other harms like relationship disruption or missing work. Much of this relevant contextual information simply cannot be collected by operators under current social norms. This missing context is why seemingly identical profiles based on available data could have different risks and, as the \citet{gamblingcommission2023risk} has noted, why high-risk players exist at all stake levels.

There are also challenges related to AI ethics generally, such as risks of bias in datasets or models \citep{percy2020lessons}, data collection protocols, and ensuring human oversight for high-consequence decisions. While a full discussion of AI ethics in gambling is beyond this paper’s scope (discussions available in \citealp{binesh2025identifying,ghaharian2025aiethics,Percy2022}), we note the trend towards translating ethical principles into regulation (e.g., the EU AI Act approved in May 2024).

\subsection{Specific Challenges for Procuring or Commissioning Models}

Those procuring or commissioning models for player risk detection face practical questions: which model performs best by a given metric (e.g., sensitivity, precision, F1 score), which metric is most important to consider, how each model performs across different player groups, and what interventions are attached to the model’s classifications. These questions are valuable for both procurement managers and those commissioning in-house tools.

Given the quantitative nature of data science, it might seem surprising that these questions are not easily addressed today. This void is often filled by competing corporate marketing claims. For example, claims about risk detection model accuracy were discussed at the International Association of Gaming Regulators annual conference \citep{percy2024trust}, with examples including claims that models can exceed 80\% accuracy \citep{sustainableinteraction}, 87\% \citep{mindwayai}, and over 90\% \citep{entain2022sustain}.

While including such numbers is a positive step for corporate transparency, these claims are nearly impossible to evaluate without critical context for at least three reasons. First, a “high accuracy” claim can be misleading. For example, in a population where only 10\% of individuals have a problem, a model could achieve 90\% accuracy simply by identifying no one as at-risk. Second, these figures are not necessarily comparable. It is not possible to conclude that a 90\%+ model is definitively better than an 87\% model, without context to ensure we are comparing like with like. Key factors such as the definition of risk, the prevalence of risk in the underlying sample, and the approach to out-of-sample testing (i.e., validating the model on new, unseen data) can all lead to dramatically different performance results, even when using the same modeling technique. Finally, these claims are not independently reviewed, and it remains unclear whether other models’ performance remains unreported for reasons of privacy or poor performance.

Reviewing published academic papers or patent filings provides additional data beyond marketing and website claims but does not provide the full detail necessary \citep{MARIONNEAU2025100644}. While additional metrics and methodological detail are available, it is rare to find exactly the same reporting criteria across studies, simply because there are so many metrics and methodological choices involved. More seriously, even where full detail is available across contexts, there are no methods to adjust resulting performance metrics for true like-for-like comparison across those different contexts. Generalizability in player risk profiles and model structures is limited across brands \citep{ghaharian2024evaluating}, across time \citep{Murch2024time}, and likely across a host of other factors such as gambling type, jurisdiction and regulatory context. As a result, it is insufficient to lean solely on better norms for consistent reporting\footnote{For examples of standardized reporting guidelines and checklists for ML models and AI-based systems, see TRIPOD-AI \citep{Collins2024}, CONSORT-AI \citep{Liu2020}, SPIRIT-AI \citep{CruzRivera2020}.}. More transparent and standardized reporting is only one step towards answering a reasonable, fundamental question from end users: \textit{How does one model’s performance truly compare to others?}

\subsection{System Consequences}

For operators, the inability to effectively compare, evaluate, or audit player risk detection systems creates a paradox. They can claim adequate RG measures without revealing specifics of their systems. However, this makes it difficult for good actors to demonstrate to regulators and the public that these measures are truly high quality. In the absence of industry standards or independent benchmarks, some operators may base decisions on cost-effectiveness and ease of integration rather than accuracy and efficacy, leading to market preference for low-cost, easily deployable solutions above model quality. This approach can cause reputational damage and incur regulatory action, especially if a major incident occurs\footnote{For example, the UK Gambling Commission fined Corbett Bookmakers ~£0.7m in March 2025, where one identified issue was “failing to identify a consumer who staked £23,674 in a 13-day period as someone who may be at risk of or experiencing harms associated with gambling” (https://www.gamblingcommission.gov.uk/news/article/gbp686-070-fine-for-corbett-bookmakers-limited).}.

There are no perfect comparative metrics, and we should recognize the limits in so-called “league table” approaches, which attempt to create an oversimplified ranking of models from best to worst. However, this does not mean comparison attempts should be shelved.

The gambling sector might reasonably look with envy at comparative metrics used in other AI-driven fields. For instance, established benchmarks for tasks like translation, image labeling, and Large Language Model (LLM) performance (e.g., intelligence, coding, agentic capabilities) are widely used\footnote{See, for example, public aggregations of these benchmarks at epoch.ai/benchmarks and artificialanalysis.ai.}.

To be clear, these existing benchmarks are not perfect and are themselves the subject of ongoing research to improve their design \citep{bean2025measuring, Hardy2024}. But they prove that useful, non-simplistic comparisons are possible. Such comparisons support sector decisions and may fuel model improvements across the board, as they have done in some AI sectors, while building confidence and trust with stakeholders. At present no such benchmarks exist for player risk detection systems, so the only way of knowing which risk model works best in a given context is to build, buy, and apply them all—a profoundly impractical option.

\section{Benchmark Datasets and Performance Benchmarking}

A fundamental practice in ML research is evaluating algorithms against standardized benchmark datasets \citep{longjohn2024benchmark, Thiyagalingam2022}. Across numerous domains, benchmark datasets have proven successful in identifying optimal ML solutions for specific problems by enabling cross-stakeholder collaboration and fostering competitive environments for innovation. This process, Performance Benchmarking, allows for objective quantitative comparisons between different approaches, as methods and results cannot be reliably compared when models are trained on different data and evaluated with different diagnostics \citep{Thiyagalingam2022}. Such standardization helps track progress over time, assess novel methods, and select the most appropriate approach for a given application. Importantly, benchmark datasets should serve as proxies for real-world tasks, so performance on the benchmark is generalizable to practical applications \citep{longjohn2024benchmark}. Beyond scientific validation, benchmarking can also serve as a critical component of regulatory infrastructure, providing a transparent and accountable framework for evaluating complex systems \citep{Percy2022}. To understand how this framework could be applied in the gambling field, it is instructive to review its successful implementation in other domains.

\subsection{Background}

One of the earliest and most well-known examples of a benchmark dataset is the Modified National Institute of Standards and Technology (MNIST) database \citep{lecun2010convolutional}. Created in 1998, MNIST consists of 70,000 size-normalized, 28x28 pixel images of handwritten digits. Each image is accompanied by a label indicating the correct number. Before MNIST, tracking progress on numeric recognition was difficult because researchers used different and incompatible datasets. MNIST provided a large, standardized collection that allowed researchers to test and benchmark their models against a consistent baseline.

Throughout the 2000s, numerous benchmark datasets emerged, applying MNIST’s fundamental principles to a growing range of use cases and more complex applications. For example, the Street View House Number (SVHN) dataset, created in 2011, contains 600,000 labeled digits cropped from Street View images \citep{netzer2011reading}, and sought to advance image recognition beyond simple handwritten digits by providing a platform for identifying characters in natural environments. Similarly, the MedMNIST benchmark was created to advance biomedical image classification in healthcare \citep{Yang2023}.

More recently, the rapid advancements of LLMs have given rise to numerous benchmarks to assess their capabilities. Many, like the Massive Multitask Language Understanding (MMLU) test\footnote{MMLU is designed to measure knowledge across diverse domains (e.g., humanities, social sciences, hard sciences) by evaluating a model’s accuracy on multiple-choice questions; for details, see its dataset card at https://huggingface.co/datasets/cais/mmlu.}, consist of large sets of multiple-choice questions across various domains \citep{https://doi.org/10.48550/arxiv.2009.03300}. Similar to how SVHN advanced image recognition, newer LLM benchmarks move beyond multiple-choice questions to assess more practical skills. For example, Vending-Bench uses a simulated environment to evaluate LLM-agents’ ability to operate a vending machine business \citep{backlund2025vending}\footnote{Vending-Bench is a simulated environment that tests how well an AI agent can manage a long-running business scenario. The agent must make decisions about inventory, orders, and pricing in order to make money (https://andonlabs.com/evals/vending-bench).}.

\subsection{Types of Benchmarking}

Benchmark datasets are as varied as ML itself and can be categorized along several axes. At a fundamental level, they can be distinguished by data type: images, audio, text, time-series, and tabular data. They can also be characterized by the ML task they address: classification, detection, regression, etc. The data curation method can also characterize benchmarks, with datasets ranging from purely synthetic data (e.g., see ACT-Thor: \citealp{hanna-etal-2022-act}) to real-world data (e.g., SVHN).

Beyond these characteristics, benchmarking can be categorized as scientific-based or competition-based. As described by \citet{dueben2022challenges}, scientific benchmarks provide a platform for addressing broad, long-term problems shared by many research groups, with performance often self-reported through peer-reviewed publications. In contrast, competition benchmarks generally focus on narrower tasks that benefit from short-term, intensive focus of a wider community. Kaggle is a well-known example where monetary awards are often given for the top-performing solution. For instance, the “American Express – Default Prediction” competition challenged participants to use anonymized transactional data to predict which customers would fail to pay their debts \citep{amex-default-prediction}. Other platforms, such as HuggingFace, Papers with Code, the UCI ML Repository, and OpenML, also facilitate benchmarking (typically without monetary components) by providing infrastructure to host data, share models, and track performance on public leaderboards. As such, the line between these categories can blur.

\subsection{The State of Benchmarking in Player Risk Detection}

Despite the success of benchmarking in other fields, its application in gambling studies remains nascent, with no formal benchmark datasets for player risk detection. The closest example is research stemming from The Transparency Project\footnote{For details on this initiative, see The Transparency Project's website: http://www.thetransparencyproject.org/}, which made several datasets publicly available. The \textit{bwin} dataset, in particular, functioned as a de facto benchmark for researchers interested in player risk detection. The \textit{bwin} dataset consists of player account records from a European-based online sports betting provider. A subset of these accounts is labeled with the reason for account closure, including “gambling problems,” which commonly serves as a proxy for problematic play. This dataset enabled foundational ML studies, from initial cluster analyses \citep{adami2013markers, braverman2012gamblers} to later classification tasks comparing various supervised ML models \citep{Philander2013}. However, the dataset’s utility is limited by its age and the fact that it represents data from only a single operator. Although ML studies have continued to emerge since The Transparency Project, these efforts remain fragmented. Research groups often produce valuable work, but it frequently relies on proprietary data that cannot be publicly shared for broader validation \citep{Ghaharian2022}.

Thus, a primary barrier to establishing benchmarks in gambling studies is data accessibility. Because player data is commercially sensitive and proprietary, it creates perceived challenges in data sharing and research transparency. Although early initiatives like The Transparency Project partially overcame these issues, their impact was limited by the technological constraints of the time. More recently, the University of Nevada, Las Vegas (UNLV) Payments Research Collaborative adopted a more modern data-sharing model, publishing a peer-reviewed paper on a dataset of gambling payment transactions and making analytic subsets available on a public repository \citep{ghaharian2023payments}. Similarly, \citet{Zendle2024} contributed by releasing a synthetic dataset\footnote{A “synthetic” dataset is an artificially generated dataset. In this case, the authors used their original dataset to generate a new dataset that mimics the statistical patterns and correlations of the original. This process allows researchers to share data for analysis without exposing any of the original, privacy-sensitive data.} derived from the banking transaction data of 424 UK-based survey respondents, linked with their Problem Gambling Severity Index (PGSI) scores. Still, without operator cooperation, these efforts are limited in their generalizability.

Regulator-coordinated requests have proven effective for targeted research: in the UK, the Gambling Commission worked with GambleAware to commission player account data across seven major operators, although barriers limited the availability of the data to third party researchers \citep{forrest2022patterns}. Additionally, legislative mandates could offer a more structural approach. For example, the Massachusetts Expanded Gaming Act requires casino licensees to provide anonymized customer tracking data to the Massachusetts Gaming Commission, which must make this data available to qualified researchers \citep{Andrews29052025}.

Beyond data accessibility, creating benchmark datasets for player risk detection presents several other complex challenges. For example, a benchmarking solution must tackle the definition of the target outcome variable, as the very concept of harm in gambling can be ambiguous, contentious, and is inherently multifaceted. Furthermore, it must be capable of integrating diverse data sources, from wagering activity to payment transactions, and account for the distinct behavioral patterns across different gambling verticals, such as sports betting, online casinos, and land-based venues. In the following Section 4, we attempt to initiate a conceptual discussion by outlining the foundational components that a robust benchmarking framework would likely require, before examining its associated challenges in Section 5. 

\section{A Proposed Benchmarking Framework for Player Risk Detection Systems}

A comprehensive benchmarking solution will require a broad suite of labeled datasets representing dimensions necessary to model risky player behaviors. The solution should provide sufficient datasets to enable industry participants to train and test risk detection models across these dimensions—we refer to this as the \textit{benchmarking suite}. For this paper we describe three core dimensions: Time, Engagement Level, and Gambling Vertical.

\subsection{Core Dimensions}

A conceptual model of the benchmarking suite outlines some core dimensions that could be captured across multiple datasets (Figure 1).

\begin{figure}[ht]
    \centering
    \includegraphics[width=0.8\textwidth]{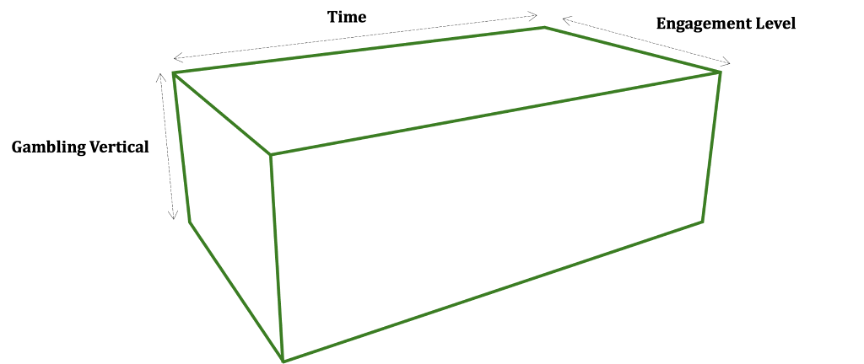}
    \caption{Core Dimensions of the Benchmarking Suite.}
    \label{fig:risk_model} 
\end{figure}

\textit{Time} refers to the length of the period of play that the data spans. For explanatory purposes in this paper, we sub-categorize time into sessions, days, and months. Providing a short-term view (i.e., sessions) is crucial for identifying “in-the-moment” risk behaviors, such as within-session loss-chasing \citep{Parke2015}. Longer-term horizons, such as days or months, are necessary to understand cumulative impacts and identify at-risk players within specific time-frames. For example, prior studies have sought to assess future risk using as little as 7 days \citep{auer2024predicting} or 30 days of data \citep{braverman2012gamblers}, while others have used 12-month datasets to create ML models \citep{kairouz2023enabling, Murch2023}.

\textit{Engagement Level} may refer to the frequency, regularity, and/or historical duration of a customer’s play. This dimension is critical as it directly impacts the nature of the data available for modeling. For instance, if we consider a dataset with a longer time horizon (e.g., 12-months) the behavioral patterns and available data for a long-term, regular player will be vastly different from those of a new or infrequent player, the latter representing a data-sparse problem with only a few sessions. From a ML perspective, it is likely that different models, features, or thresholds are required to effectively detect risk in these distinct cohorts. Additionally, ML models deployed without regard to engagement levels might do little more than distinguish those who play regularly from infrequent gamblers, instead of identifying those patterns that are specifically associated with gambling problems or harms among regular gamblers. Thus, a robust benchmarking suite would need to contain datasets that represent the range of engagement levels to ensure player risk detection systems can be assessed across varying engagement levels.

\textit{Gambling Vertical} refers to the type of gambling products offered. The main gambling verticals by market share are lottery, casino, and sports betting. Within these verticals there are sub-categories—for example, table games, slots, and live dealer in casino, draw products and instant win games in lottery, and fixed odds, prop, and parlay in sports betting. While ML models can be developed to encapsulate player activity across gambling verticals, prior work has suggested superior performance when vertical-specific models are developed (e.g., see \citealp{kairouz2023enabling}). Thus, a benchmarking suite should strive to provide datasets that can be used to develop both broad and vertical-specific models. 

\subsection{Example Benchmark Datasets}

To illustrate the capabilities and practicality of the benchmarking suite concept, we present four possible benchmark datasets that may be provided (Figure 2). These examples are not intended to be exhaustive; rather, we offer them as a conceptual starting point, representing some of the most common or likely scenarios. Each example demonstrates how the framework’s dimensions can provide different, measurable evaluation scenarios.

For each of these examples we make the assumption that each benchmark dataset consists of players’ gambling data (i.e., session and/or transaction level information), is sufficient in size, and has the required labels to build ML and predictive models (e.g., players labeled as at-risk or not). We detail specific data requirements later in this section. Additionally, while acknowledging the presence of other verticals, our examples and illustrations cover the top three verticals (i.e., lottery, casino, sports betting).

Importantly, a single benchmark dataset can be used to support multiple benchmark tasks. A task is defined as the specific ML exercise a participant must complete. For instance, a dataset containing PGSI scores as the target variable could support multiple tasks. These might include a binary classification problem, where the goal is predicting whether a player is at risk based on PGSI scores (e.g., 0 = no risk, 5+ = at-risk), or a multi-class classification problem, where the goal is predicting risk across multiple PGSI buckets (e.g., 0, 1-2, 3-4, 5-7, 8+). However, acknowledging that validated self-report measures are not always available, tasks could also be defined around other harm proxies (See Section 5.3 for a discussion).

\begin{figure}[ht]
    \centering
    \includegraphics[width=0.8\textwidth]{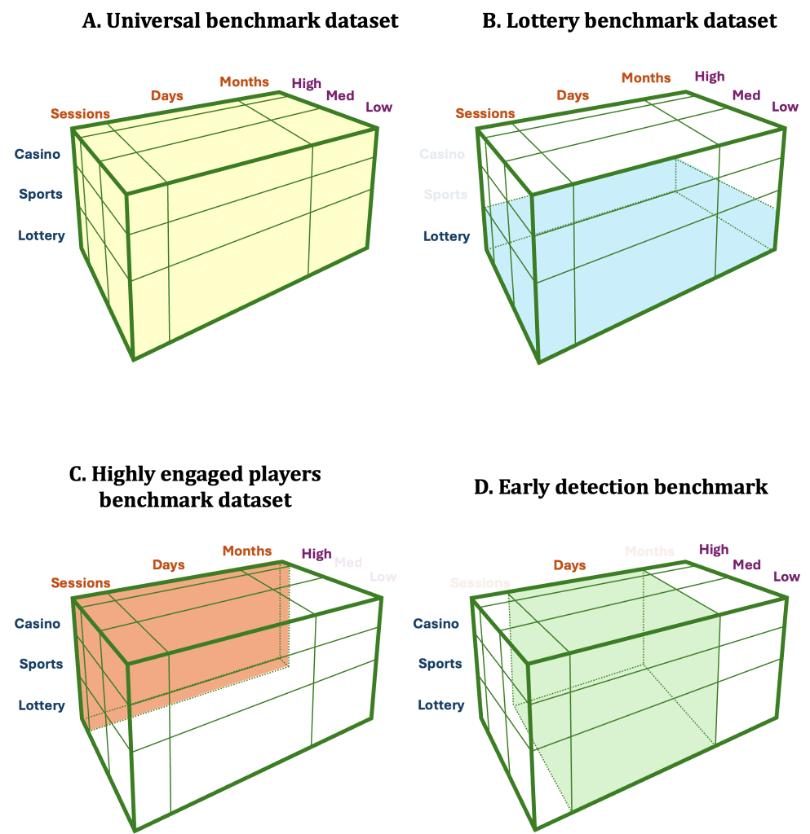}
    \caption{Examples of Benchmark Datasets.}
    \label{fig:risk_model2} 
\end{figure}

A. \textit{Universal Benchmark Dataset}: If an operator is multi-vertical (i.e., offers a wide range of gambling products), and has built one risk detection algorithm to be used across its entire player database, that operator can test that algorithm on a benchmark dataset that encompasses player activity across each of the core dimensions. In other words, this labeled dataset would contain session-level activity records that can be aggregated to longer time horizons for players with various engagement levels across a range of gambling products, providing variability on the Time, Engagement Level, and Gambling Vertical dimensions.

B. \textit{Lottery Players Benchmark Dataset}: Some operators focus on offering a particular gambling vertical, for example lottery, and therefore would benefit from a benchmark on this subset of players. Additionally, evidence suggests that ML models for player risk detection may work best when they are tuned for specific activities (Ghaharian et al., 2023; Kairouz et al., 2023). 

C. \textit{Highly Engaged Players Benchmark Dataset}: Data scientists typically focus on sub-populations where data is more abundant and build models around these populations, therefore a useful benchmark for industry could be more highly engaged players who have reasonably long tenure (e.g., months rather than days). 

D. \textit{Early Detection Benchmark Dataset}: It is critical to detect risk as early as possible to enable timely interventions and prevent future harm. Therefore, this benchmark dataset would specifically assess an algorithm’s ability to predict future harm using only short-term data. It would use data from a player’s initial days of activity (e.g., days 1-7) to predict a negative outcome much further in the future (e.g., a risk flag at 6 months). Unlike the previous benchmark (C), this assessment could cover all engagement levels, not just the highly-engaged subset.

While the preceding examples (B–D) focus on isolating a single core dimension, real-world benchmark datasets will likely cross multiple dimensions simultaneously, which may better reflect the complex nature of player risk detection in practice. For instance, detecting risky player sessions (e.g., those involving frequent top-ups or failed deposits) on slot products would require a dataset spanning two dimensions: Gambling Vertical (e.g., slots) and Time (sessions). Similarly, a common and valuable benchmark for a casino operator might combine the Gambling Vertical and Engagement Level dimensions, specifically to assess algorithms on more highly engaged, high-value players. Ultimately, these types of composite benchmark datasets are vital for assessing algorithms under conditions that are representative of the actual operational environment.

\subsection{Requirements}

As \citet{dueben2022challenges} describe, “Benchmark datasets are defined by the community. While many datasets can be proclaimed as ‘benchmarks,’ they are only successful if the ML community embraces them.” Similarly, the success of a benchmarking suite for player risk detection hinges on its adoption by relevant stakeholders. As a foundation for future implementation of such a benchmarking suite, we present initial discussion on data specifications as well as various technical and usability requirements. Many of the administrative requirements could be satisfied by existing data repositories (for in-depth discussion see: \citealp{longjohn2024benchmark}) and could be worked into a customized repository specific for this task domain.

\subsubsection{Data collection}

Data collection involves the curation of data into the benchmarking suite. It is unlikely that the breadth of global stakeholders in the gambling sector will collectively cooperate to curate a single, comprehensive dataset representing the entire gambling industry. Likewise, the practicality, in terms of complexity and cost, of curating such a global dataset is likely to be significant. A more realistic scenario involves gradual curation and procurement of data from various stakeholders over time. In the past, industry stakeholders have donated data to academic institutions (e.g., The Transparency Project and UNLV Payments Research Collaborative), regulators have required licensees to transfer data to centralized storage for audit purposes\footnote{For example, the Netherlands Gambling Authority (Kansspelautoriteit, Ksa) mandates the use of a Controle Data Bank (CDB).}, and there are examples of industry consortia supporting efforts to share data across operators\footnote{For example, the Responsible Online Gaming Association (ROGA) has announced the creation of a data clearing house that will consolidate and share player self-exclusion data among its members to advance consumer protection. Similarly, the GamProtect scheme facilitated by the Betting and Gaming Council seeks to create a mechanism for participating operators to compliantly and securely share information with each other.}.

\subsubsection{Dataset documentation}

As data are curated, they must be presented in a way that allows stakeholders (including academics, researchers, and industry operators) to effectively use these datasets for evaluating ML models.

Because raw data are rarely ready for immediate use, they require processing to become a usable benchmark dataset. The benchmarking suite must facilitate recording of any processing steps taken, and given the sensitivity of player protection – where model performance directly impacts human well-being – there should also be an attempt to enforce quality review. As highlighted by \citet{longjohn2024benchmark}, there is currently a lack of quality review of datasets, and existing repositories often view themselves as simply providing infrastructure to support data sharing rather than providing review or quality assurance processes. To address this, the proposed benchmarking suite could require the publication of peer-reviewed introductory papers to accompany datasets, similar to those found in Data in Brief (e.g., see: \citealp{ghaharian2023payments}).

The entire suite would house numerous datasets. Thus, users must understand each dataset’s context and structure. Best practices from established data science platforms and repositories can provide a strong foundation (e.g., HuggingFace, UCI ML Repository, OpenML). For instance, datasets should include core metadata such as the size and shape of the data, list a creator/owner of the dataset (who can act as a point of contact for queries and/or issues), as well as versioning and timestamps. Some datasets may well reside in raw format (e.g., session-level activity for each bet placed), whereas others may represent aggregated information (e.g., session-level activity rolled up to daily or weekly information about players’ activity). Critically, documentation should specify the temporal relationship between the player activity data and any associated labels (the target variable). This includes defining the timeframe the activity data covers (e.g., the first 7 days of play, 12-months since registration date) and its temporal proximity to the target variable (e.g., a PGSI score collected at 6 months, the date self-exclusion occurred). Tasks associated with benchmark datasets could also be listed alongside this metadata and/or with further details about each task (e.g., train-test splits, primary evaluation metric, etc.) provided in a separate location within the repository (e.g., as a downloadable file or a link to a dedicated page). We provide an example of such metadata for the Early Detection Benchmark Dataset (example D above) in Table 1.

\begin{table}[H]
    \centering
    \caption{Example Metadata for a Benchmark Dataset}
    \label{tab:validation_options}
    \vspace{3mm}
    \begin{tabular}{|l|>{\raggedright\arraybackslash}p{0.75\linewidth}|}\hline
         \textbf{Attribute}& \textbf{Description}\\\hline
         Dataset name& The Matrix Casino: Early Risk Dataset (Zion)\\\hline
         Description& Contains session-level data from The Matrix online casino (Zion, Earth). The dataset includes 105,677 sessions from 5,465 unique players during their first 7 days of activity.\\\hline
         Dimensions& Time: 7 days \newline Engagement level: All players who completed PGSI \newline Vertical: Casino\\\hline
         Benchmark task(s)& \textbf{\underline{B1 7 day detection}}: Supervised Classification. To predict a player’s future harm status using only their initial 7 days of session data. \newline \textbf{\underline{B2 1 day detection}}: Supervised Classification. To predict a player’s future harm status using only their initial day of session data.\\\hline
         Target variable& Risk\_Flag: A binary flag (0 or 1) indicating high-risk status at 6 months, defined by a PGSI score of 5+.\\\hline
         Size& 105,677 rows (sessions)\\\hline
         Data fields& Player\_ID: Unique player identifier.\newline
Transaction\_Type: A deposit or withdrawal of real money.\newline
Transaction\_Amount: Amount of real money deposited or withdrawn.\newline 
Transaction\_Method: Method of payment: card, PayPal etc.\newline
Transaction\_Status: Status of the transaction: approved, declined, etc.\newline
Start\_Time: Session start timestamp (UTC).\newline
End\_Time: Session end timestamp (UTC).\newline
Bet\_Count: Number of wagers in the session.\newline
Total\_Staked: Total monetary value staked.\newline
Net\_Outcome: Net win/loss from the session.\newline
Product: Gambling product used.\newline
Risk\_Flag: (Target) Player’s risk status at 6 months.
\\\hline
         Data dictionary& Download\_link\\\hline
         Creator \/ Author& Mr. Thomas A. Anderson \\\hline
         Citation& Anderson, T. A. (2025). The Matrix Casino: Early Risk Dataset (Zion) https://doi.org/00.0000/prd.bs.2025.1117\\\hline
         Versioning& V.1 2025-11-17\\ \hline
    \end{tabular}
    
\end{table}

This “data work” is a substantial undertaking but is essential to the downstream success of any benchmark dataset. Unfortunately, this type of work is often undervalued and under-incentivized \citep{gero2023incentive}. To incentivize and recognize these efforts, datasets could be treated as citable intellectual contributions through persistent identifiers, such as a Digital Object Identifier (DOI) \citep{longjohn2024benchmark}.

\subsubsection{Dataset access and usability}

Datasets and associated benchmark tasks should be easy to access and use. Datasets should be provided in common, accessible formats (e.g., CSV, various application programming interfaces [API] formats) with detailed documentation on how to access data.

For tasks, a baseline model or script (e.g., logistic regression or random forest) could be provided in a popular language (e.g., Python). This serves as a reference implementation, demonstrating how to load the data and generate predictions. It also establishes a performance baseline that more sophisticated models can attempt to surpass, offering an immediate point of comparison.

\subsubsection{Result submission and validation} 

Benchmark tasks will also require a trustworthy mechanism for users to submit their model’s results and have them validated. To ensure fair comparisons, users must adhere to a clear set of rules. There are several potential approaches for managing this process, each with its own trade-offs, we outline these in Table 2.

\begin{table}[H]
    \centering
    \caption{Result Submission and Validation Options for Benchmark Tasks}
    \label{tab:validation_options2}
    \vspace{3mm}
    \begin{tabular}{|>{\raggedright\arraybackslash}p{0.2\linewidth}|>{\raggedright\arraybackslash}p{0.4\linewidth}|>{\raggedright\arraybackslash}p{0.3\linewidth}|}\hline
         \textbf{Method}&  \textbf{Description}& \textbf{Strengths and limitations}
\\\hline
         Publication-based verification&  The solution can act as a data repository (e.g., UCI Machine Learning Repository) and results are validated indirectly via academic peer review.& Rigorous academic peer review but slow and not standardized.
\\\hline
         Full code and model submission&  Participants submit complete source code, configuration, and model details along with their results, enabling the benchmark owners to directly reproduce outcomes.& This is maximum reproducibility, but conflicts with potential IP concerns.
\\\hline
         Prediction submission&  Akin to Kaggle, users run models locally on the benchmark dataset, then submit only the predictions which are scored against hidden labels.& This addresses IP concerns, but cannot know how predictions were generated.
\\\hline
         Container submission&  Users submit an executable model (e.g., via a Docker container) and the benchmark owners run the model against a hidden test set.& Addresses not knowing how predictions were generated, but requires more engineering and resources.
\\ \hline
    \end{tabular}
    
    \vspace{1ex} 
    \parbox{0.9\linewidth}{
        \small \textit{Note:} The tasks could potentially support all submission pathways and implement a “badge system” to recognize different levels of validation (e.g., bronze for prediction-only, silver for containerized submission, gold for full code + publication).
    }
\end{table}

\section{Challenges and Considerations}

While our framework in Section 4 provides a foundational structure, it is intended as a conceptual guide. We recognize that the practical implementation of such a suite faces significant hurdles and limitations not fully addressed in this initial proposal. This section, therefore, discusses several key challenges and practical considerations that our field must navigate to move from this conceptual model to a functional, robust benchmarking system.

\subsection{Governance and Administration}

A central question in implementing such a framework is: \textit{who should govern and administer the system?} This governing entity would be responsible for overseeing all functions including data collection, administration of contributors and users, ensuring correct documentation, facilitating usability, and validating the robustness and trustworthiness of the benchmark process and its results.

Administration by a governmental body is likely to improve coordination by operators and overcome collective action problems inherent in a competitive private market. Regulators can legally mandate the submission of standardized data from all licensed operators, ensuring the comprehensiveness and consistency required for a foundational dataset. 

A third-party model involving a neutral entity like an international research institute or data trust presents a distinct advantage in this regard \citep{Paprica2020}. Such an organization is not constrained by jurisdictional borders and could therefore aggregate data across multiple states, provinces, and nations. This transnational approach would enable a far more comprehensive dataset.

\subsection{Dataset Size}

A robust benchmark dataset must address sample considerations that extend beyond conventional calculations of statistical power.  These include challenges with representativeness and detecting low-prevalence phenomena. For instance, problematic gambling, as identified by a proxy like the PGSI, has a low base rate, necessitating a very large overall sample just to capture enough positive cases for reliable model training and evaluation. Without an adequate number of these cases, a model’s ability to accurately identify at-risk players becomes statistically unreliable.

Furthermore, the database must accommodate a wide range of models, from simple logistic regression to complex deep learning networks. Since more complex models require larger datasets to prove their effectiveness, sample sizes should be large enough for the most sophisticated systems. While sourcing such a dataset could prove challenging, potential solutions include the construction of synthetic datasets, government-mandated data aggregation, or the pooling of data through a third party.

The use of synthetic data provides a potentially constructive path to scale dataset size while avoiding issues with player privacy and commercial data sharing sensitivity \citep{samsel2025}. By generating an artificial dataset that replicates the statistical properties of real-world data without containing any personally identifiable information, synthetic data allows for broad access for research and model testing while the source data remains secure. The viability of this approach, however, is entirely contingent upon the quality and representativeness of the original dataset from which the synthetic version is derived, as well as the assumptions of the approach used to generate the synthetic data \citep{Chen2021,Susser2024}.  

\subsection{Defining the Target Variable}

A review of the academic literature and existing solutions reveals at least three different categories of outcome targets for ML models in player risk detection.

First, proxy indicators of harm such as account closure and, more commonly, voluntary self-exclusion (VSE) are used by both industry and academic studies (e.g., \citealp{finkenwirth2021using,hopfgartner2023predicting,Percy2016}). These are routinely collected alongside the data used to make predictions, can easily delineate the at-risk group through binary classification (VSE or no VSE), and are not subject to self-report bias or measurement error. However, their reliability as indicators of at-risk status is questionable. Players can choose to self-exclude or close their accounts for various reasons beyond the experience of gambling problems, including to pre-empt harms or to close one’s online account \citep{catania2021understanding,auer2016should}. One strategy to mitigate this risk is to use a subset of self-excluders who have a sufficiently long track record of play with the operator, to ensure players who self-exclude relatively quickly after opening an account are not used for model training.

Second, screening instruments such as the PGSI \citep{ferris2001canadian} or Brief Biosocial Gambling Screen \citep{gebauer2010optimizing} have been used as target variables in recent academic studies \citep{auer2023using,hopfgartner2024using,kairouz2023enabling,Murch2023,Perrot2022} and by Focal Research Consultants \citep{hancock2008gambling,Schellinck2011}. Self-report measures offer a more reliable indicator of risk status, although are more difficult to obtain and link with the data used to make predictions, are subject to self-report biases (particularly if administered by the operator), and require a threshold for classification as at-risk to be selected in advance \citep{Murch2023}. Additionally, the measures reflect a respondent’s overall gambling behavior, which may span multiple platforms, operators, or even jurisdictions, limiting its interpretability as a label for data from any single operator’s platform. Furthermore, because self-reported risk status can change over time in response to variations in gambling engagement, repeated administration of screening instruments would improve datasets. Longitudinal measurement of the target variable would allow models to be trained and evaluated on their ability to predict transitions into or out of at-risk status, rather than relying solely on cross-sectional classification.

Third, a final approach may be the use of single indicators such as loss chasing or substantial variations from normal play, which are themselves sometimes used to classify consumers as at-risk in the absence of ML \citep{delfabbro_behavioural_2024,McAuliffe2022}. Several researchers have attempted to develop thresholds for standard behavioral indicators (e.g., frequency of gambling, amount spent) that differentiate lower- from higher-risk gambling to form lower risk gambling limits \citep{hodgins2023lower,louderback2021developing} – these, or similar threshold-style indicators, could plausibly serve as targets, either individually or in combination.

Overall, the variation and lack of a standard approach to classification presents a challenge to constructing and comparing ML models, as there is no agreed-upon specification to label training data appropriately. Given the diversity of outcome variables and definitions of risk, we do not propose a single ground truth. Instead, we recommend a suite of benchmarks that can evolve over time as the domain continues to develop and as a clearer definition of “at-risk” emerges. We do, however, advocate for transparency in how at-risk status is determined to ensure the process can be replicated in real-world environments and to help acknowledge the biases inherent in different approaches. For example, human-based labeling of cases based on opaque reasoning would not satisfy this criterion, but labels determined using a clear and reproducible set of steps would.

\subsection{Demographic and Jurisdictional Diversity}

Another challenge is significant diversity in player populations and jurisdictions. A successful benchmarking framework must avoid over-representing datasets from one demographic group (e.g., all men) or market (e.g., North America), except where the benchmark results are explicitly only applied within those settings. Over-representing certain demographic groups in benchmark datasets risks validating the performance of models that perform poorly in real-world settings with more diverse populations and could potentially favor biased models that disproportionately target or ignore minority groups (see \citealp{Murch2024}).

There is evidence that ML models can effectively detect at-risk gambling in datasets from countries not included in training data, although training on local data can improve performance \citep{hopfgartner2024using}. A fairer benchmarking framework will offer datasets from a variety of regions to allow contributors to demonstrate the performance of their model(s) in preferred settings (e.g., EU vs. North America).

\subsection{Dataset Characteristics}

Jurisdictional variation may also impact the ease with which models are applied due to differences in feature availability. Additionally, differences in available products and other policies across jurisdictions will affect model applicability. This may skew the composition and relevance of benchmark datasets developed in those contexts. For example, a model built using online tracking data in Australia where players can cancel their withdrawals may attribute high importance to this behavior, which would not be present in datasets from jurisdictions that have prohibited this feature (e.g., Great Britain; \citealp{gamblingcommission2021consultation}). This highlights the need for clear documentation that describes the characteristics of each benchmarking dataset and idiosyncrasies resulting from regulation or product innovation.

Those experienced in working with tracking data will be familiar with variations in labeling (e.g., “stake” vs. “amount”; “bonus stake” vs. “free stake”) and the operationalization of variables (e.g., winning bets recorded as negative vs. positive values; aggregation or separation of multi-bet legs) between datasets from different operators. While minor, these discrepancies could preclude the efficient application of models. For benchmarking to work, a harmonized data dictionary and standardized preprocessing steps will be essential to produce benchmark datasets that others can replicate in model training.

A further consideration is the temporal structure of the data. Academic studies report building models using varying timeframes of tracking data, ranging from 12 months (e.g., \citealp{Murch2023,Perrot2022}) to just 30 days \citep{hopfgartner2024using}. As discussed in Section 4, there is a need to provide benchmarks for different periods that span months to days. Ideally, benchmark datasets will span an extended period (e.g., 12 months) to permit varied testing on the same dataset and the use of diverse timeframe aggregations (e.g., monthly or daily behavioral averages). A benchmark for acute risk detection (e.g., 1 to 7 days) may benefit from uniquely large datasets, since these can include all active and newly registered customers, not only those with extensive historical data.

\subsection{The Evolving Gambling Landscape and need for Dynamic Benchmarks}

A final challenge is the dynamic and evolving nature of gambling. Models built to accurately classify or predict stable outcomes such as handwriting or images can be continuously evaluated against benchmark datasets with little need for updating. By contrast, gambling environments and consumer behavior can be continuously impacted by changes in regulation, local economies, and product innovation, which may result in data drift. Validation of a model against a historical dataset (such as those added to The Transparency Project) may hold little weight when evaluating contemporary performance. As such, ML benchmarks for gambling cannot be static. Government mandates that require data sharing (see \citealp{Newall2024}) are likely to be the most successful way of accessing periodically updated datasets for a benchmarking suite. Regardless, successfully sharing and deploying the first iteration of benchmarking datasets without compromising privacy will be essential to building trust with operators and the public alike, and ensuring continuous buy-in from industry stakeholders.

\section{Conclusion}

Technology, specifically AI and its subsets such as ML, are being increasingly used to support industry efforts to minimize harms. While we applaud efforts to leverage technology to better protect consumers, the current state of affairs is sub-optimal, in that it is not possible to credibly ascertain the efficacy of solutions that use ML to detect at-risk play. As AI becomes increasingly integrated into diverse processes, ranging from business to health and education, benchmarks are becoming increasingly developed to enable more objective assessment and comparison of their efficacy for the task at hand.

The gambling industry has an opportunity to leverage these technological developments and lead in responsible and ethical AI by promoting greater transparency and trust in the harm detection solutions they develop and deploy. Benchmarking is key to this, and there are significant examples across other use cases and industries on how benchmarking can be developed successfully.

The concept of benchmarking is particularly important in any gambling industry practice that relies on ML models to support decision making in the areas of consumer protection, fairness, and crime. For example, these concepts could be applied to assessing the efficacy of sports betting integrity (i.e., match-fixing) detection algorithms, anti-money laundering (AML) detection algorithms, and impacts of operator customer RG interventions.

While we outline a broad range of challenges and considerations, we argue that these are not insurmountable and can be addressed with the right combination of data access, domain and technical expertise, and the ability to develop trustworthy solutions. A failure of industry to address these concerns will cast doubt on the efficacy of these solutions and increase pressure on regulators to take greater control of how they are developed and used. In Spain, the Directorate General for the Regulation of Gambling (DGOJ) recently received a legal mandate to establish a unified ML model that will be mandatory for all licensed operators in Spain \citep{directorate2025}. Such regulatory developments are likely to continue expanding if the transparency of existing industry methods remains opaque.


\bibliographystyle{elsarticle-harv}
\bibliography{references} 

\section*{ORCID iDs}
\noindent
Kasra Ghaharian: \url{https://orcid.org/0000-0003-4238-0278}\\
Simo Dragicevic: \url{https://orcid.org/0009-0005-1905-787X}\\
Chris Percy: \url{https://orcid.org/0000-0003-0574-9160}\\
Sarah E. Nelson: \url{https://orcid.org/0000-0001-7967-4910}\\
W. Spencer Murch: \url{https://orcid.org/0000-0003-2780-3578}\\
Robert M. Heirene: \url{https://orcid.org/0000-0002-5508-7102}\\
Kahlil Simeon-Rose: \url{https://orcid.org/0000-0002-0747-0772}

\section*{Disclosures}

During the last five years, International Gaming Institute (IGI) at University of Nevada, Las Vegas, has received funding for its research and programs from Action Gaming, American Gaming Association, Aristocrat Leisure Limited, Association of Gaming Equipment Manufacturers, Axes.ai, Bet Blocker, Clarion Gaming, DraftKings, Entain Foundation, ESPN, Evoke plc, Focal Research Consultants, Gaming Analytics, Global Payments, IGT, Kindbridge Behavioral Health, Las Vegas Sands Corporation, Massachusetts Gaming Commission, MGM Resorts International, Playtech plc, Responsible Online Gaming Association, Yuhaaviatam of San Manuel Nation, Sightline Payments, Sports Betting Alliance, State of Nevada Department of Health and Human Services, State of Nevada Knowledge Fund, Walker Digital Table Systems, and Wynn Resorts Ltd. Additionally, IGI organizes the triennial International Conference on Gambling and Risk Taking, a research-focused event supported by sponsors from industry, academia, and the legal/regulatory sectors; a full list of sponsors for the most recent conference can be found at \url{https://www.unlv.edu/igi/conference/18th/sponsors}. IGI is home to an industry-focused advisory board (\url{https://www.unlv.edu/igi/advisory-board}), and specific programs, such as AiR Hub, have their own advisory panels. These advisory roles include resource support, and individual advisors are required to adhere to IGI policies. IGI maintains a strict research policy (\url{https://www.unlv.edu/igi/research-policy}), as well as a partnership and transparency framework (\url{https://www.unlv.edu/igi/policies/partnership}), to ensure appropriate firewalls exist between funding entities and IGI’s research and programs.

During the past 5 years, Kasra Ghaharian has received funding for research and/or consulting services from the Nevada Department of Health and Human Services, the Nevada Governor’s Office of Economic Development, the Massachusetts Gaming Commission, AXES.ai, Playtech, Sightline, IGT, Differential, Focal Research Consultants, GP Consulting, and the International Center for Responsible Gaming. Ghaharian has received honoraria/travel reimbursement from the Responsible Gambling Council, the Illinois Council on Problem Gambling, and Kindred Group. None of these entities played roles in the design, analysis, or interpretation of research, and imposed no constraints on publishing.

During the past 5 years Simo Dragicevic has received funding for employment, research and consulting services from The European Lotteries, The Gambling Commission (Great Britain), Playtech Plc, and Yaspa. Simo is also Adjunct Fellow at the International Gaming Institute (IGI). 

Chris Percy provides consulting services to the gambling industry, with recent activity including research for the UK Gambling Commission and the BetBuddy player risk detection software and responsible gambling R\&D at Playtech Plc.

During the past five years, Sarah E. Nelson has served as a paid grant reviewer for the New Frontiers in Research Fund (Government of Canada), the International Center for Responsible Gaming (ICRG), and the National Institutes of Health (NIH). She has also received travel reimbursement and speaker honoraria from the ICRG, Responsible Gaming Association of New Mexico, Responsibility.org, and the Maryland Highway Safety Office. She received publication royalty fees from the American Psychological Association, and received course royalty fees from the Harvard Medical School Department of Continuing Education.

W. Spencer Murch holds a postdoctoral fellowship from the Alberta Gambling Research Institute, a joint initiative by the Alberta Provincial Government, and the Universities of Calgary, Lethbridge, and Alberta. WSM holds a research grant from the International Centre for Responsible Gaming (USA) that did not fund his contributions to this research. WSM served on the Advisory Board on Safer Gambling, an expert group that formerly advised the UK Gambling Commission.

Robert M. Heirene's contribution to this work was supported by a post-doctoral fellowship from the New South Wales Responsible Gambling Fund as part of the Gambling Research Capacity Grant program, administered by the Office of Responsible Gambling. RMH has worked on a project funded by Responsible Wagering Australia (a representative body of Australian online wagering operators; University of Sydney, 2019–2021) and as an independent, sub-contracted statistical consultant for PRET Solutions Inc on a commissioned project (funded by the Australian Casino operator Crown; 2023). RMH has also received research funding from the International Centre for Responsible Gaming (ICRG) and the Brain and Mind Center, University of Sydney.

Khalil Simeon-Rose declares the following financial interests/personal relationships which may be considered as potential competing interests: In the three-year period from 2022 to 2025, KS received research funding from the Cambridge Health Alliance, Division on Addiction; he received consulting payments from Eilers \& Krejcik Gaming, Scientific Affairs, VictorStrategies, MGM Resorts, and the United Arab Emirates’ General Commercial Gaming Regulatory Authority; and he received expert witness payments from Benjamin Leigh Carter. KS holds unpaid appointments on the advisory board of the Responsible Online Gaming Association, the Canadian Gaming Association’s responsible gaming sub-committee, and the Deutsche Stiftung Glücksspielforschung gGmbH Scientific Advisory Board.

Tracy Schrans declares that, during the past 5 years, Focal Research Consultants have received funding from various gaming companies, regulatory bodies, government funding agencies, and research institutions for consultancy work in the area of player data analytics, social responsibility, and safer gambling products, services, and compliance solutions. None of these entities played a role in the design, analysis, or interpretation of this study, and imposed no constraints on publishing.

\end{document}